\documentclass[12pt,preprint]{aastex}

\begin{document}

\title{First bounds on the very high energy $\gamma$-ray emission from
Arp 220}

\author{
 J.~Albert\altaffilmark{a}, 
 E.~Aliu\altaffilmark{b}, 
 H.~Anderhub\altaffilmark{c}, 
 P.~Antoranz\altaffilmark{d}, 
 A.~Armada\altaffilmark{b}, 
 C.~Baixeras\altaffilmark{e}, 
 J.~A.~Barrio\altaffilmark{d},
 H.~Bartko\altaffilmark{g}, 
 D.~Bastieri\altaffilmark{h}, 
 J.~Becker\altaffilmark{f}
% ...
}
 \altaffiltext{a} {Universit\"at W\"urzburg, D-97074 W\"urzburg, Germany}
 \altaffiltext{b} {Institut de F\'\i sica d'Altes Energies, Edifici Cn., E-08193 Bellaterra (Barcelona), Spain}
 \altaffiltext{c} {ETH Zurich, CH-8093 Switzerland}
 \altaffiltext{d} {Universidad Complutense, E-28040 Madrid, Spain}
 \altaffiltext{e} {Universitat Aut\`onoma de Barcelona, E-08193 Bellaterra, Spain}
 \altaffiltext{f} {Universit\"at Dortmund, D-44227 Dortmund, Germany}
 \altaffiltext{g} {Max-Planck-Institut f\"ur Physik, D-80805 M\"unchen, Germany}
 \altaffiltext{h} {Universit\`a di Padova and INFN, I-35131 Padova, Italy} 
% ...

\begin{abstract}
Using the Major Atmospheric Gamma Imaging Cherenkov Telescope
(MAGIC), we have observed the nearest ultra-luminous infrared galaxy
Arp 220 for about 15 hours. No significant signal was detected
within the dedicated amount of observation time. The first upper
limits to the very high energy $\gamma$-ray flux of Arp 220 are
herein reported and compared with theoretical expectations.
\end{abstract}


\section{Introduction}

The large masses of dense interstellar gas and the enhanced number
densities of supernova remnants and massive young stars present in
starburst galaxies suggest that they might emit $\gamma$-ray
luminosities orders of magnitude greater than normal galaxies and be
related to the production of cosmic rays (see, e.g., Torres et al.
2004, Torres \& Anchordoqui 2005). Of the taxonomy of starbursts, a
place of privilege is given to luminous and ultraluminous infrared
galaxies (luminous and ultra-luminous IR galaxies (LIRGs and ULIRGs,
defined as having log$(L_ {\rm IR}/L_\odot) > 11$ and 12,
respectively, see Sanders \& Mirabel (1996) for a review about these
objects). Such star-forming environments emit a large amount of
infrared (IR) radiation, because of the abundant dust molecules
reprocessing of stellar UV photons.

Therefore, the infrared luminosity, $L_{\rm IR}$, of a galaxy can
(but not always) be an indication of star formation taking place in
it.  LIRGs are the dominant population of extragalactic objects in
the local universe ($z<0.3$) at bolometric luminosities above $L >
10^{11}$ L$_\odot$, and ULIRGs are in fact the most luminous local
objects. Most ULIRGs appear to be recent galaxy mergers in which
much of the gas of the colliding objects has fallen into a common
center (typically less than 1 kpc in extent), triggering a huge
starburst (e.g., Sanders et al. 1988, Melnick \& Mirabel 1990). The
size of the inner regions of ULIRGs, where most of the gas is found,
can be as small as a few hundreds parsecs, and populated with dense
molecular environments (e.g., Gao \& Solomon 2003a, 2003b) that
makes them prone to have large star formation events and cosmic-ray
densities. However, no LIRG, nor ULIRG, nor any other starburst
galaxy has been detected in $\gamma$-rays, not even at the EGRET
energy range above 100 MeV. Upper limits were imposed for M82 and
NGC~253, the two nearest starbursts, as well as for many LIRGs
(Torres et. al. 2004, Cillis et al. 2005). At higher energies, HESS
has recently reported upper limits for NGC 253 (Aharonian et al.
2005) and we are aware of no limit reported yet for Arp 220.

Arp 220 is the nearest ULIRG (located at about 72 Mpc) and the best
studied. A complete multiwavelength modeling from radio to TeV
$\gamma$-rays was presented by Torres (2004), where an extensive
description on the observational knowledge on this object can also
be found. Arp 220 possess the record as to being the object with the
highest directly measured supernova explosion rate known, with
recent measurements placing it at an outrageous 4 $\pm$ 2 per year
(Lonsdale et al. 2006). Such a high supernova explosion rate
emphasizes the quality of Arp 220 as a possible $\gamma$-ray target.

\section{Observations}

MAGIC (see e.g., Baixeras et al. 2004, Cortina et al. 2005) for a
detailed description) is a single dish Imaging Air Cherenkov
Telescope. Located on the Canary Island La Palma ($28.8^\circ$N,
$17.8^\circ$W, 2200~m a.s.l.), the telescope has a 17-m diameter
mirror, and it is equipped with a 576-pixel $3.5^\circ$
field-of-view photomultiplier (PMT) camera. The analogue PMT signals
are transported via optical fibers to the trigger electronics and
are read out by a 300 MSamples/s FADC system. MAGIC's angular
resolution is approximately 0.1$^{\circ}$, energy resolution is
about 20\%, and the trigger (analysis) threshold is 55 (90) GeV.

Arp 220 celestial coordinates are: (J2000) $\alpha=15^{\rm
h}~34^{\rm m}~57\fs21$, $\delta=+23\degr~30\arcmin~09\farcs$5, what
have allowed MAGIC to observe it always with zenith angle lower than
20 degrees. Observations proceeded from May to June, 2005, for a
total time of 947 min, of which 925 min where selected -after
quality checks- for further analysis. MAGIC observations were
carried out in the ON-OFF mode, with a similar amount of ON and OFF
data being considered for the analysis. This observation mode allows
a reliable background estimation. The specific properties of the
used data sample, both ON and OFF, can be found in
Domingo-Santamar\'ia (2006), where additional details on the quality
checks performed and on the cut optimization using an independent
Crab sample of data are also available.

The data analysis was carried out using the standard MAGIC analysis
and reconstruction software (Bretz \& Wagner 2003), the first step
of which involves the calibration of the raw data (Gaug et~al.
2005). It follows the general stream presented in (Albert et al.
2006a,b,c): After calibration, image cleaning tail cuts of 10
photoelectrons (phe) for image core pixels and 5 phe (boundary
pixels) have been applied (see e.g. Fegan 1997). These tail cuts are
accordingly scaled for the larger size of the outer pixels of the
MAGIC camera. The camera images are parameterized by image
parameters (Hillas 1985). In this analysis, the Random Forest method
(see Bock et~al. 2004, Breiman 2001) for a detailed description) was
applied for the $\gamma$/hadron separation.

The source position-independent image parameters  SIZE, WIDTH,
LENGTH, CONC (Hillas 1985) and the third moment of the phe
distribution along the major image axis were selected to
parameterize the shower images. After the training, the Random
Forest method allows to calculate for every event a parameter, the
HADRONNESS, which is a measure of the probability that the event is
not $\gamma$-like. The $\gamma$-sample is defined by selecting
showers with a HADRONNESS below a specified value, which is
optimized using a sample of Crab data which has been processed with
the same analysis stream. An independent sample of Monte Carlo
$\gamma$-showers was used to determine the cut efficiency.

\begin{figure}
\centering \vspace{1cm}
\includegraphics[width=.75\textwidth]{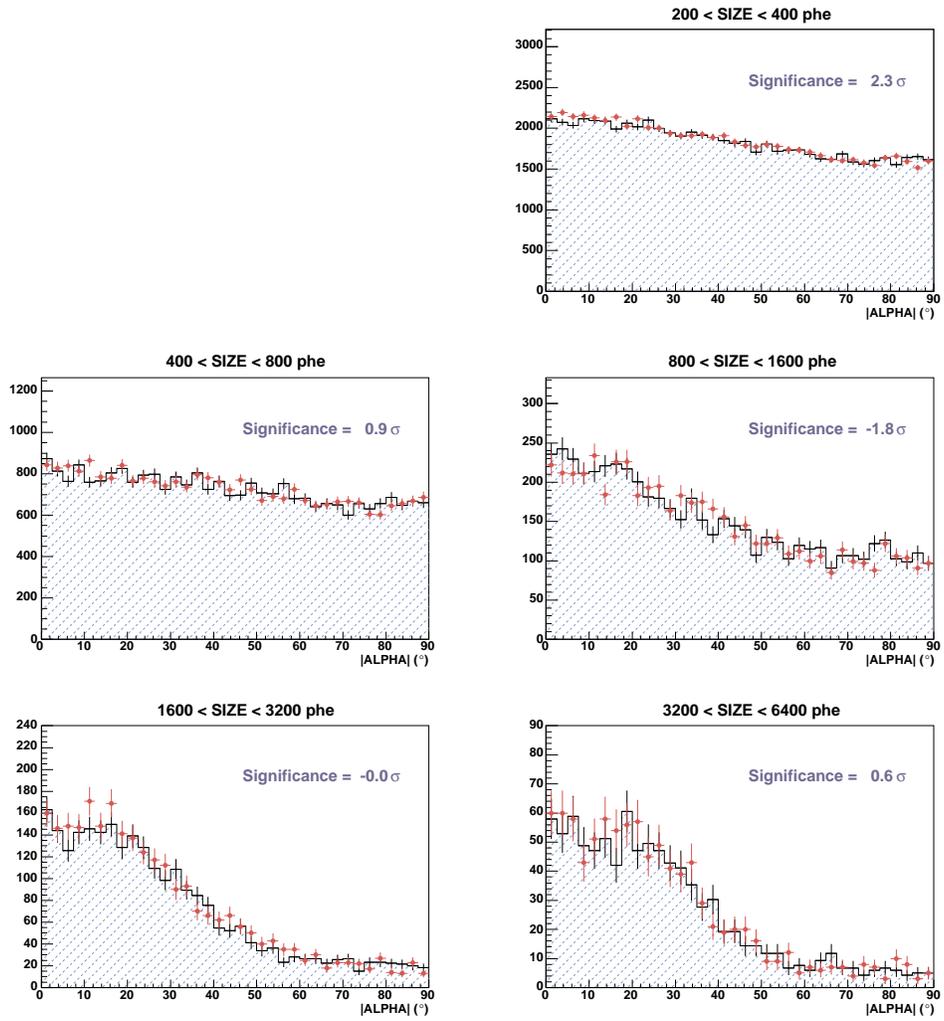} 
\caption{ALPHA plots for the Arp 220 data, separated in bins of
SIZE.} \label{arp-alphaPlots}
\end{figure}

Figure \ref{arp-alphaPlots} shows the ALPHA plots which were
obtained for the whole sample of ON and OFF data, after applying the
HADRONNESS cuts. The ON and OFF ALPHA distributions match reasonably
well within fluctuations. No signal above the background level is
observed in any of the SIZE bins. Details are given in Table
\ref{significance-Arp-table}. There, we also give the Crab rate
(obtained with 1.22 hours of data under the same analysis stream
applied to Arp 220) . The upper limits to the very high energy
$\gamma$-ray flux from Arp 220 can then be calculated out of the
number of excess and background events, in a similar way as that
used in the case of LS I +61 303 (Albert et al. 2006d), using the
method by Rolke and Lopez (2001). The final column of Table
\ref{significance-Arp-table} summarizes these results.

\section{Discussion and Concluding Remarks}

\begin{figure}
\centering
\includegraphics[width=.5\textwidth]{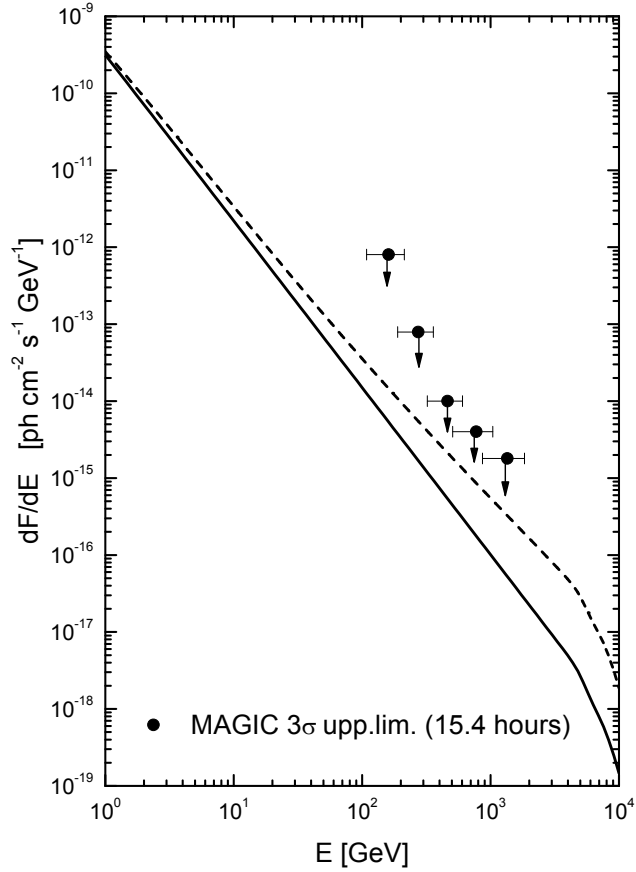}
\caption{MAGIC upper limits to the differential $\gamma$-ray flux of
Arp 220. The curves represent the theoretical predictions, a solid
line shows the result using the $\delta$-function (Aharonian \&
Atoyan 2000) or Kamae et al. (2005) approximations for the
proton-proton cross section, whereas a dashed line shows the result
using the parameterization proposed by Blattnig et al. (2000),
extrapolated to high energies. Theoretical curves are from (Torres
2004, and Torres \& Domingo-Santamar\'ia 2005).} \label{upper-lim}
\end{figure}

\begin{deluxetable}{ccccccccc}
\tabletypesize{\scriptsize} \rotate \tablecaption{Number of excess
and background events and the corresponding significance and upper
limits obtained from the Arp 220 analysis.} \tablewidth{0pt}

\tablehead{ \colhead{SIZE bin} & \colhead{Avg. energy} &
\colhead{HADR.} & \colhead{ALPHA} & \colhead{\# excess } &
\colhead{\# bkg.} & \colhead{N$_{\sigma}$} & \colhead{Crab rate} &
\colhead{3$\sigma$ Upp.Limit    } \\  \colhead{(phe) } & \colhead{
(GeV)  } & \colhead{cut} & \colhead{cut ($^{\circ}$)} &
\colhead{events} & \colhead{events} &  \colhead{($\sigma$)}  &
\colhead{($\gamma$/hr)} & \colhead{ (10$^{-14}$ ph cm$^{-2}$
s$^{-1}$ GeV$^{-1}$)     } }

 \startdata

200 -- 400  &   160.7 &   0.20   &      10.00         &    295    &   8343    &    2.3        &  171.3            &    80.14          \\ 
 400 -- 800  &   273.6 &   0.20   &       8.75         &     51    &   2873    &    0.7        &  182.8            &    7.89           \\ 
 800 -- 1600 &   462.7 &   0.16   &       7.50         &    -62    &    707    &   $-$1.8      &  90.2             &    1.00           \\ 
1600 -- 3200 &   773.3 &   0.20   &       6.25         &      7    &    370    &    0.3        &  81.1             &    0.40           \\ 
3200 -- 6400 &   1351.8 &   0.18   &       6.25         &     22    &    132    &    1.4        &  33.6            &    0.18           \\ 

\enddata

 \label{significance-Arp-table}
\end{deluxetable}

The upper limits imposed to the differential flux of Arp 220 with
15.4 hours of data are above the theoretical curves at all
energies.\footnote{The theoretical curves are the result of a
multiwavelength modeling discussed by Torres (2004), and Torres \&
Domingo-Santamar\'ia (2005). From the proton steady state population
the computation of the secondary $e^\pm$  proceeds considering
knock-on interactions and decay of charged pions. The lepton
population is let to evolve to its steady state, computing the
energy losses by synchrotron emission, ionization, bremsstrahlung,
inverse Compton, and adiabatic expansion, and their confinement
timescale. The radio spectrum is evaluated from the steady electrons
synchrotron emission, modulated by free-free absorption.  Infrared
emission from dust is also simulated to describe the observational
data. These photons are the seed of inverse Compton process. Once
the multiwavelength spectrum of the object has been reproduced, the
high energy $\gamma$-ray emission is evaluated through the decay of
neutral pions, and bremsstrahlung and inverse Compton of the steady
electron population. Photon absorption is considered to obtain the
final predictions of fluxes.} All upper limits are at least about
one order of magnitude above the curve obtained using the
$\delta$-function (Aharonian \& Atoyan 2000) or Kamae et al. (2005)
approaches for the proton-proton cross section parameterization. The
latter (see the appendix in Domingo-Santamar\'ia \& Torres 2005 for
a detailed discussion) are the most reliable parameterizations of
the proton-proton cross section, which is devoid of the intrinsic
problems of extrapolating Blattnig et al.'s (2000) formulae to high
energies. The current results imply that under the same conditions
of telescope performance and sensitivity of the data analysis
applied, the amount of observation time needed to be devoted to Arp
220 in order to be at the level of confirming or rejecting the
predictions from the $\gamma$-ray emission of this object is too
high for a detector with a typical duty cycle of about 1000 hours
per year, such as MAGIC. Disregarding the fact that Arp 220 is the
nearest ULIRG and the galaxy with the largest supernova explosion
rate we know, its distance dilutes the putative $\gamma$-ray flux it
produces. MAGIC upper limits are consistent with this interpretation
and with the theoretical prediction that results from a complete
multiwavelength modeling of the object. The first detection of
$\gamma$-rays from starburst regions beyond our galaxy is yet to be
achieved.

\section*{Acknowledgments}

We would like to thank the IAC for the excellent working conditions
at the Observatory de los Muchachos in La Palma. The support of the
German BMBF and MPG, the Italian INFN and the Spanish CICYT is
gratefully acknowledged. This work was also supported by ETH
Research Grant TH~34/04~3 and the Polish MNiI Grant 1P03D01028.

\end{document}